\documentclass[reprint,amsmath,amssymb,nofootinbib,aps,pra]{revtex4-2}

\usepackage{graphicx}
\usepackage{dcolumn}
\usepackage{bm}
\usepackage[breaklinks=true,colorlinks=true,linkcolor=blue,urlcolor=blue,citecolor=blue]{hyperref}
\usepackage{multirow}
\usepackage[dvipsnames]{xcolor}
\usepackage[pscoord]{eso-pic}
\usepackage[normalem]{ulem}
\usepackage{slashed}
\usepackage{makecell}

\newcommand{\tquote}[1]{``#1''}

\begin{document}
	
\title{Tensions in the dark: shedding light on Dark Matter-Dark Energy interactions}

\author{Matteo Lucca}
\email{mlucca@ulb.ac.be}
\author{Deanna C. Hooper}
\email{deanna.hooper@ulb.be}
\affiliation{Service de Physique Th\'{e}orique, Universit\'{e} Libre de Bruxelles, C.P. 225, B-1050 Brussels, Belgium}

\begin{abstract}
	The emergence of an increasingly strong tension between the Hubble rate inferred from early- and late-time observations has reinvigorated interest in nonstandard scenarios, with the aim of reconciling these measurements. One such model involves interactions between Dark Matter and Dark Energy. Here we consider a specific form of the coupling between these two fluids proportional to the Dark Energy energy density, which has been studied extensively in the literature and claimed to substantially alleviate the Hubble tension. We complement the work already discussed in several previous analyses and show that, once all relevant cosmological probes are included simultaneously, the value of the Hubble parameter in this model is $H_0=69.82_{-0.76}^{+0.63}$~km/(s Mpc), which reduces the Hubble tension to $2.5\sigma$. Furthermore, we also perform a statistical model comparison, finding a $\Delta\chi^2$ of $-2.15$ (corresponding to a significance of 1.5$\sigma$) with the inclusion of one additional free parameter, showing no clear preference for this model with respect to $\Lambda$CDM, which is further confirmed with an analysis of the Bayes ratio.
\end{abstract}

\maketitle

\section{Introduction}
Despite the remarkable success of the standard $\Lambda$CDM model across many different scales, recent advances in precision cosmology have yielded  discrepancies in observations at different redshifts (see e.g. \cite{Freedman:2017yms} for a nice historical overview), which have opened the door to several alternative models.

The most striking examples of so-called tensions between the $\Lambda$CDM predictions inferred form the latest analysis of the \textit{Planck} Collaboration \cite{Aghanim:2018eyx} and independent observational sources involve the expansion rate of the universe, quantified with the Hubble parameter $H_0$, and the value of $\sigma_8$, a quantity reflecting the amount of late-time matter clustering. The former has been measured by the Hubble Space Telescope (HST) using local distance ladder measurements, with the SH0ES Collaboration reporting ${H_0=74.03\pm1.42~\text{km/(s Mpc)}}$ \cite{Riess:2019cxk}, which is at a 4.4$\sigma$ difference to the value measured by the \textit{Planck} satellite, ${H_0=67.4\pm0.5~\text{km/(s Mpc)}}$ \cite{Aghanim:2018eyx}. Furthermore, for the latter quantity cosmic shear surveys such as the Kilo Degree Survey (KiDS) \cite{Koehlinger2017} or the Dark Energy Survey (DES) \cite{Troxel:2017xyo, Abbott:2017wau, Krause:2017ekm} have been employed, leading to tensions in the $\sigma_8$ measurements of the order of 2-3$\sigma$.

Many different models have been proposed to solve these tensions (for a non-exhaustive list see e.g. \cite{Lancaster:2017ksf,Oldengott:2017fhy,DiValentino:2017oaw,Kazantzidis:2018rnb,Kreisch:2019yzn,Escudero:2019gvw,Park:2019ibn,Archidiacono:2016kkh,Lesgourgues:2015wza,Buen-Abad:2017gxg,Archidiacono:2019wdp,Poulin:2016nat,Binder:2017lkj,Bringmann:2018jpr,Hooper:2019gtx,Poulin:2018cxd,Agrawal:2019lmo,Lin:2019qug,Pan:2019gop,Desmond:2019ygn,Kazantzidis:2019dvk}), including models which question the nature of Dark Matter (DM) or Dark Energy (DE). For instance, one particularly promising and well-studied model consists in interacting DM and DE (henceforth iDMDE, see e.g. \cite{Wang_2016} for a recent review). 

Although the first studies of iDMDE appeared in the 1990s \cite{Casas_1992, Wetterich:1994bg, Anderson:1997un, Amendola:1999er, Bean2001, Farrar_2004}, this class of theories saw renewed interest roughly a decade ago, with the first computations of the cosmological perturbation equations \cite{Valiviita_2008, He:2008si, He:2009mz, Gavela:2009cy, Gavela:2010tm, Honorez:2010rr}. Furthermore, in recent years additional effort has been dedicated to the evaluation of cosmological constraints on iDMDE \cite{Pettorino:2013oxa, Ade:2015rim, Xia:2016vnp, Murgia:2016ccp, Yang:2016evp, Pan2017, An:2017kqu, Kumar:2017dnp, Santos:2017bqm, Guo:2017deu, Pan:2017ent, An:2017crg, DiValentino:2017iww, Wang:2018azy, vonMarttens:2018iav, Yang:2018qec, Costa:2019uvk, Martinelli:2019dau, Li:2019loh, Kumar:2019wfs, DiValentino:2019ffd, DiValentino:2019jae, Yang:2020zuk, Gomez-Valent:2020mqn}. Most recently, these models have also been tested using the gravitational-wave observations from Ligo and Virgo \cite{Yang:2019vni, Bachega:2019fki, Li:2019ajo}.

However, as already argued in e.g. \cite{Wang_2016} (see in particular Sec.~2 therein), the choice made to describe the energy transfer term $Q$ between DM and DE is -- to a large extent -- arbitrary. In fact, from a quantum field theory perspective, the Lagrangian defining the interaction between the fermionic DM field $\psi$ and a quintessential field $\varphi$ reads \cite{Farrar_2004, Wang_2016} 
\begin{equation}\label{eq: Lag}
\mathcal{L}=\frac{1}{2} \partial^{\mu} \varphi \partial_{\mu} \varphi-V(\varphi)+i \bar{\psi}
\slashed{\partial} \psi+M(\varphi) \bar{\psi} \psi\,,
\end{equation}
where $V(\varphi)$ is the scalar field potential and $M(\varphi)$ is a time-varying mass term which describes the effective interaction between the two fields. However, although some attempts have been made to justify a particular form of the coupling term \cite{Yin2007, Costa2015, DAmico:2016jbm, DAmico:2018hgc, Pan:2020zza}, in most cases the definition of $M(\varphi)$ has only been assumed to be linearly \cite{Anderson:1997un, Farrar_2004, Baldi2011, Wang_2016} or exponentially \cite{COMELLI2003115} dependent on $\varphi$. The same arbitrariness is also common in the choice of the potential $V(\varphi)$, which can have either a power-law \cite{Lucchin1985, WETTERICH1988668, Ferreira1998, Baldi2011b} or an exponential \cite{COMELLI2003115} behavior, or a combination of the two~\cite{BRAX199940, Baldi2011b}. One can then show that the value of $Q$ in the cosmological conservation equations is a function of $V(\varphi)$ and $M(\varphi)$ \cite{Wang_2016}, and thus inherits the same justification problems.

In cosmological contexts, in order to compensate the missing derivation of $Q$ from first principles, one often intuitively assumes that it depends on the energy densities involved, i.e. $\rho_c$ (for DM) and $\rho_x$ (for DE), and on the expansion rate $H$. Given this freedom, a variety of possible interactions have been considered in the literature (see e.g. Sec.~2 of \cite{Wang_2016} and \cite{Pan:2020zza} for more complete discussions). Following the steps of the recent analysis by \cite{DiValentino:2019ffd, DiValentino:2019jae}, in this work we limit our selection to one single model that has gained great popularity due to its potential impact on the $H_0$ tension. In this model, the interactions between DM and DE over cosmological scales are ruled by a term linearly proportional to the DE energy density. Note that a similar dependence on, for instance, the DM energy density has been shown to be unstable for couplings larger than approximately $10^{-2}$, disfavoring this form of interactions \cite{Valiviita_2008, He:2009mz, Kumar:2016zpg, Pan:2020zza}.

In \cite{DiValentino:2019ffd, DiValentino:2019jae} it was shown that when using current data, such as temperature, polarization, and lensing data from \textit{Planck}, as well as the supernovae measurements from HST, this iDMDE model can considerably alleviate the $H_0$ tension, although not fully solve it. Furthermore, the aforementioned papers found a preference for a nonzero value of the interactions when using the combination of \textit{Planck} and the most recent SH0ES data.

However, it has also been shown in the literature (see e.g.~\cite{Bernal:2016gxb, Poulin:2018zxs, Knox:2019rjx}) that models relying on late-time modifications to the expansion history (aiming to increase $H_0$ today and reconcile the tension) are not compatible with the combination of Baryon Acoustic Oscillation (BAO) and Supernovae Type Ia data. Indeed, this combination of data sets probes the low-redshift expansion history, providing a model-independent constraint on the product $H_0 r_s$, where $r_s$ is the sound horizon. As $r_s$ is only directly sensitive to prerecombination physics, any model that tries to increase $H_0$ after recombination will not be able to satisfy the constraint on $H_0 r_s$, leading to a \emph{no-go theorem} for such models.

Nonetheless, the iDMDE model studied here is not a subcase of the DE models (of the form $\Lambda$CDM+$(w_0,w_a)$) studied in~\cite{Bernal:2016gxb, Knox:2019rjx}, as it has an interacting DM component that introduces differences at both the background and perturbation level. As such, the conclusions of the aforementioned papers may not apply to this model. Furthermore, while BAO and Pantheon data have been included in previous analyses (see e.g.~\cite{DiValentino:2019jae}), these have not been included simultaneously, and thus the constraint on the product $H_0 r_s$ has not been exploited. With this in mind, here we aim to test if the iDMDE model is also subject to the \emph{no-go theorem}, and thus not a viable solution to the $H_0$ tension.

This paper is organized as follows. First in Sec.~\ref{sec: math} we briefly review the theory describing the iDMDE model, and revisit the different formalisms encountered in the literature. In Sec.~\ref{sec: meth} we discuss the method and different data sets we use to evaluate this model. In Sec.~\ref{sec: res} we first reproduce part of the results of \cite{DiValentino:2019ffd, DiValentino:2019jae}, and then extend these results with complementary -- and insightful -- combinations of current data sets. A final summary of this work and additional discussions are given in Sec. \ref{sec: conc}.

\section{The mathematical setup}\label{sec: math}
We have implemented the mathematical structure describing iDMDE in the Boltzmann code \texttt{CLASS} \cite{blas2011cosmic} (version 2.7.2).

In this work we investigate a very well-studied parametrization of the energy transfer function between the DM and DE fluids \cite{Gavela:2009cy, Gavela:2010tm, Honorez:2010rr, He:2010im, Salvatelli:2013wra, Murgia:2016ccp, DiValentino:2017iww, Kumar:2017dnp, Kumar:2019wfs, DiValentino:2019ffd, DiValentino:2019jae}, which can be expressed in the 4-component notation as
\begin{align}\label{eq: Q}
Q^{\nu}=\xi H \rho_x u^{\nu}_{c}\,,
\end{align}
where $\xi$ is the coupling constant and $u^{\nu}_{c}$ is the DM 4-velocity. Although this form of the DM-DE interaction cannot be derived from Lagrangians such as the one expressed in Eq. \eqref{eq: Lag}, possible phenomenological derivations have been discussed e.g., in \cite{Pan:2020zza}. As in most of these references, we also choose $Q^{\nu}$ to be parallel to $u^{\nu}_{c}$, which avoids momentum transfer in the DM rest frame and circumvents fifth force constraints.

At the background level, the only modifications to the $\Lambda$CDM model are due to the fact that the DM and DE energy densities are not conserved singularly any more, but instead are coupled via the energy transfer $Q$, leading to
\begin{align}
\label{eq: cons_eqs 1} \dot{\rho}_{c}+3 \mathcal{H} \rho_{c} &=Q\,, \\ \dot{\rho}_{x}+3 \mathcal{H}(1+w) \label{eq: cons_eqs 2} \rho_{x} &=-Q\,,
\end{align}
where the index $c$ refers to cold DM, the index $x$ to DE, and $w$ is the DE equation of state (EOS) parameter. For our choice of $Q$, Eqs.~\eqref{eq: cons_eqs 1}-\eqref{eq: cons_eqs 2} can be analytically solved to find
\begin{align}
& \rho_{c}=\rho_{c, 0} a^{-3}+\xi\frac{\rho_{x, 0}a^{-3}}{3w_{x}^{\rm eff}}\left[1-a^{-3w_{x}^{\rm eff}}\right]\,, \\
& \rho_{x}=\rho_{x, 0} a^{-3(1+w_{x}^{\rm eff})}\,,
\end{align}
where $w_c=0$ is implicitly assumed for the DM EOS, and we have introduced
\begin{equation}
w_{x}^{\rm eff} =w+\frac{Q}{3 \mathcal{H} \rho_{x}}
\end{equation}
following \cite{Gavela:2009cy, Honorez:2010rr}.
\begin{table*}
	\centering
	\begin{tabular}{  c@{\hskip 0.8 cm}  c@{\hskip  0.8 cm}  c@{\hskip  0.8 cm}  c@{\hskip  0.8 cm}  c  }
		\hline\rule{0pt}{3.0ex} 
		Parameter & \textit{Planck} & \thead{\textit{Planck} \\ + R19} & \thead{\textit{Planck} \\ + BAO \\ + Pantheon} & \thead{\textit{Planck} \\ + R19 \\ + BAO \\ + Pantheon} \\[0.1 cm]
		\hline\rule{0pt}{3.0ex} 
		$\omega_{\rm cdm}$           & $0.059_{-0.018}^{+0.017}$ & $0.043_{-0.020}^{+0.021}$ & $0.1099_{-0.0037}^{+0.0093}$ & $0.0990_{-0.0081}^{+0.011}$ \\[0.1 cm]
		$H_0 \, [\text{km/(s Mpc)}]$ & $ 72.7_{-3.2}^{+2.4}$     & $74.0_{-1.3}^{+1.4}$      & $68.78_{-0.74}^{+0.54}$ & $69.82_{-0.76}^{+0.63}$     \\[0.1 cm]
		$\xi$                        & $-0.45_{-0.33}^{+0.16}$   & $-0.56_{-0.14}^{+0.13}$   & $ >-0.22 $ & $-0.179_{-0.074}^{+0.090}$  \\[0.1 cm]
		\hline\rule{0pt}{3.0ex}
		$\Delta\chi^2$               & $-3.60$                   & $-17.58$                  & $-0.14$          & $-2.15$  \\[0.1 cm]
		$\sigma$            		 & $ 1.9 $		   		 	 & $ 4.2 $			 		 & $ 0.4 $  		&  $ 1.5$  \\[0.1 cm]
		$2 \ln \mathcal{B}$          & $ 2.4 $		   			 & $ -14.2 $				 & $ 3.7 $  		&  $-1.1 $ \\[0.1 cm]
		\hline
	\end{tabular}	
	\caption{Mean and 68\% C.L. of the parameters most significantly affected by the presence of iDMDE (the lower bound is given at the 95\% C.L. instead), for different data set combinations. Additionally, we show three different statistical analyses of iDMDE compared to the $\Lambda$CDM model: the $\Delta\chi^2$, $\sigma$, and $2 \ln \mathcal{B} $ (as explained in the text).}
	\label{tab:bestfits}
\end{table*}

The other modification to the $\Lambda$CDM model is at the perturbation level. In the synchronous gauge, one obtains \cite{Gavela:2010tm, Honorez:2010rr, Salvatelli:2013wra, DiValentino:2017iww, DiValentino:2019ffd, DiValentino:2019jae}
\begin{align}
\label{eq: pert_1} \dot{\delta}_{c}= &-\theta_{c}-\frac{\dot{h}}{2}\left(1-\frac{\xi}{3} \frac{\rho_{x}}{\rho_{c}}\right) +\xi \mathcal{H} \frac{\rho_{x}}{\rho_{c}}\left(\delta_{x}-\delta_{c}\right)\,, \\ 
\label{eq: pert_3} \dot{\theta}_{c}=&-\mathcal{H} \theta_{c}\,, 
\\ \nonumber \dot{\delta}_{x}=&-(1+w)\left[\theta_{x}+\frac{\dot{h}}{2}\left(1+\frac{\xi}{3(1+w)}\right)\right]+ \\ & \hspace{0 cm} -3 \mathcal{H}(1-w)\left[\delta_{x}+\frac{\mathcal{H} \theta_{x}}{k^{2}}(3(1+w)+\xi)\right]\,, \\ 
\label{eq: pert_2} \dot{\theta}_{x}=& 2 \mathcal{H} \theta_{x}\left[1+\frac{\xi}{1+w} \left(1-\frac{\theta_{c}}{2\theta_{x}}\right)\right]+\frac{k^{2}}{1+w} \delta_{x}\,,
\end{align}
with initial conditions for the DE perturbations given by \cite{Salvatelli:2013wra, DiValentino:2017iww}
\begin{align}
\delta_{x}^{in}(x) =(1+w-2 \xi)C  \hspace{0.4 cm}\text{and}\hspace{0.4 cm} \theta_{x}^{in} =k^2\tau C\,,
\end{align}
where
\begin{equation}
C=-\frac{1+w+\xi / 3}{12 w^{2}-2 w-3 w \xi+7 \xi-14}\frac{2 \delta_{\gamma}^{in}}{1+w_{\gamma}}\,.
\end{equation}
In the above expression $\delta_{\gamma}^{in}= \delta_{\gamma}^{in}(k,\tau)$ are the initial conditions for the photon density perturbations, and $w_\gamma=1/3$ is the photon EOS parameter. Here we have neglected the center of mass velocity for the total fluid, $v_T$ in \cite{Gavela:2010tm, DiValentino:2017iww, DiValentino:2019ffd, DiValentino:2019jae}. Additionally, the DE sound speed has been set to unity, i.e., $c_{s,x}^2=1$, while for the DE adiabatic sound speed we have $c_{a,x}^2=w$ (see e.g. Sec.~2.3 of \cite{Valiviita_2008} for more details). 

Moreover, it is interesting to notice that for the same model, \cite{He:2008si, He:2010im, Costa:2013sva, Murgia:2016ccp} employed a different set of equations compared to Eqs.~\eqref{eq: pert_1}-\eqref{eq: pert_2}. Although the analytical derivation of both sets of equations is beyond the scope of this work, we have cross-checked that the two formulations lead to the same results (a quantitative comparison will not be discussed further within this work but can be found in e.g. \cite{Costa2015}). For sake of transparency and completeness, a version of \texttt{CLASS} including both possible sets of the perturbation equations has been made publicly available\footnote{\url{https://github.com/luccamatteo/class_iDMDE}}. 

\section{Method and cosmological probes}\label{sec: meth}
We have performed Markov Chain Monte Carlo (MCMC) scans on the iDMDE model presented in Sec. \ref{sec: math} using the parameter inference code \texttt{MontePython} \cite{audren2013conservative, Brinckmann:2018cvx} (version 3.2.0). We have judged the MCMCs to be converged using the Gelman-Rubin convergence criterion, requiring $|R-1|<0.01$ for all parameters~\cite{gelman1992}.

In the choice of priors for the initial parameters, particular care has been devoted to the DE EOS parameter $w$ and the coupling constant $\xi$. In fact, it is clear from Eqs.~\eqref{eq: pert_1}-\eqref{eq: pert_2} that $w=-1$ would create divergences. Furthermore, \cite{Gavela:2009cy} pointed out that the value of the coupling has to have opposite sign with respect to $w+1$, i.e. for $w+1>0$ one has $\xi<0$ (and vice versa), in order to avoid early-time instabilities. For these reasons, we set $w=-0.999$, consistent with the literature \cite{Salvatelli:2013wra, DiValentino:2017iww, DiValentino:2019ffd}, since this value is close enough to $-1$ to recover $\Lambda$CDM if $\xi=0$ and avoids the gravitational instabilities occurring at $w=-1$ at the same time. Note that, although the same result could have been achieved with $w=-1.001$, previous studies including $w$ as free parameter suggest a solution of the type $w>-1$ \cite{DiValentino:2017iww} (see in particular the case including also BAO and the joint light-curve analysis of the reference). As a consequence of this choice, we impose a negative value for $\xi$ as a prior. While extensive analysis allowing $w$ to vary as an additional free parameter can be found in e.g. \cite{Murgia:2016ccp, DiValentino:2019jae}, we will however not explore this avenue further within this work.

With these considerations we end up with a 6+1 extension of the standard $\Lambda$CDM model including 
\begin{equation}
\{h, \omega_{\rm b}, \omega_{\rm cdm}, n_s, \ln(10^{10}A_s), \tau_{\rm reio} \}+\xi\,.
\end{equation}
In order to constrain this set of parameters, we base our analyses on the combination of several cosmological probes. 

First of all, we consider Cosmic Microwave Background (CMB) temperature, polarization, and lensing constraints from \textit{Planck} 2018~\cite{Aghanim:2018eyx}, making use of the \textit{Planck} baseline (high-$\ell$ TT,TE,EE + low-$\ell$ EE + low-$\ell$ TT + \textit{Planck} lensing\footnote{Note that our choice of including the lensing likelihood as part of the \textit{Planck} baseline is justified by the compatibility of these liklihoods, as shown in Table 2 of \cite{DiValentino:2019jae}. We have confirmed  that the incluision of the lensing likelihood does not significantly modify the bounds on the resulting cosmological parameters presented in Table~\ref{tab:bestfits}.}, referred to henceforth as \textit{Planck}). In order to test the ability of this model to solve the $H_0$ tension, we will additionally include a Gaussian prior of the form ${H_0=74.03\pm1.42}$~km/(s Mpc), as reported by the SH0ES Collaboration~\cite{Riess:2019cxk} (referred to henceforth as R19), and also done in \cite{DiValentino:2019ffd}. Additionally, we will include the Pantheon data \cite{Scolnic:2017caz}, which contains distance moduli information of 1048 Supernovae Type Ia. Moreover, in this work we also investigate the constraining power of BAO data, using measurements of $D_V/r_{\rm drag}$ by 6dFGS at $z = 0.106$~\cite{Beutler:2011hx}, by SDSS from the MGS galaxy sample at $z = 0.15$~\cite{Ross:2014qpa}, and additionally by BOSS from the CMASS and LOWZ galaxy samples of SDSS-III DR12 at $z = 0.2 - 0.75$~\cite{Alam:2016hwk} (referred to henceforth as BAO). A similar set of probes has already been considered in \cite{DiValentino:2019jae}.

Finally, in order to determine which model is preferred by the data we will make use of three different statistical tools. First, we use a simple $\Delta \chi^2$ comparison, which allows us to break down the individual contribution from each data set. Second, we use the significance $\sigma$, which additionally takes into consideration the degrees of freedom in the different models. Third, we use the Bayes ratio, which further takes into consideration the priors of the models. We define the Bayes ratio as 
\begin{equation}
\mathcal{B} = \frac{\mathcal{E}(\mathcal{D}|\mathrm{iDMDE})}{\mathcal{E}(\mathcal{D}|\Lambda\mathrm{CDM})}\, ,
\label{eq:bayes}
\end{equation}
where $\mathcal{E}(\mathcal{D}|\mathrm{M})$ is the evidence of a model $M$ given the data $\mathcal{D}$. With this, following the Jeffrey's scale as modified by Kass and Raftery~\cite{Kass1995}, a negative (positive) value of $2 \ln \mathcal{B}$ indicates a preference for iDMDE ($\Lambda$CDM). To compute the evidence from our MCMC chains, we use the numerical code \texttt{MCEvidence}~\cite{Heavens:2017afc}.

\section{Results}\label{sec: res}
\begin{figure}[t]
	\centering
	\includegraphics[width=0.8\columnwidth]{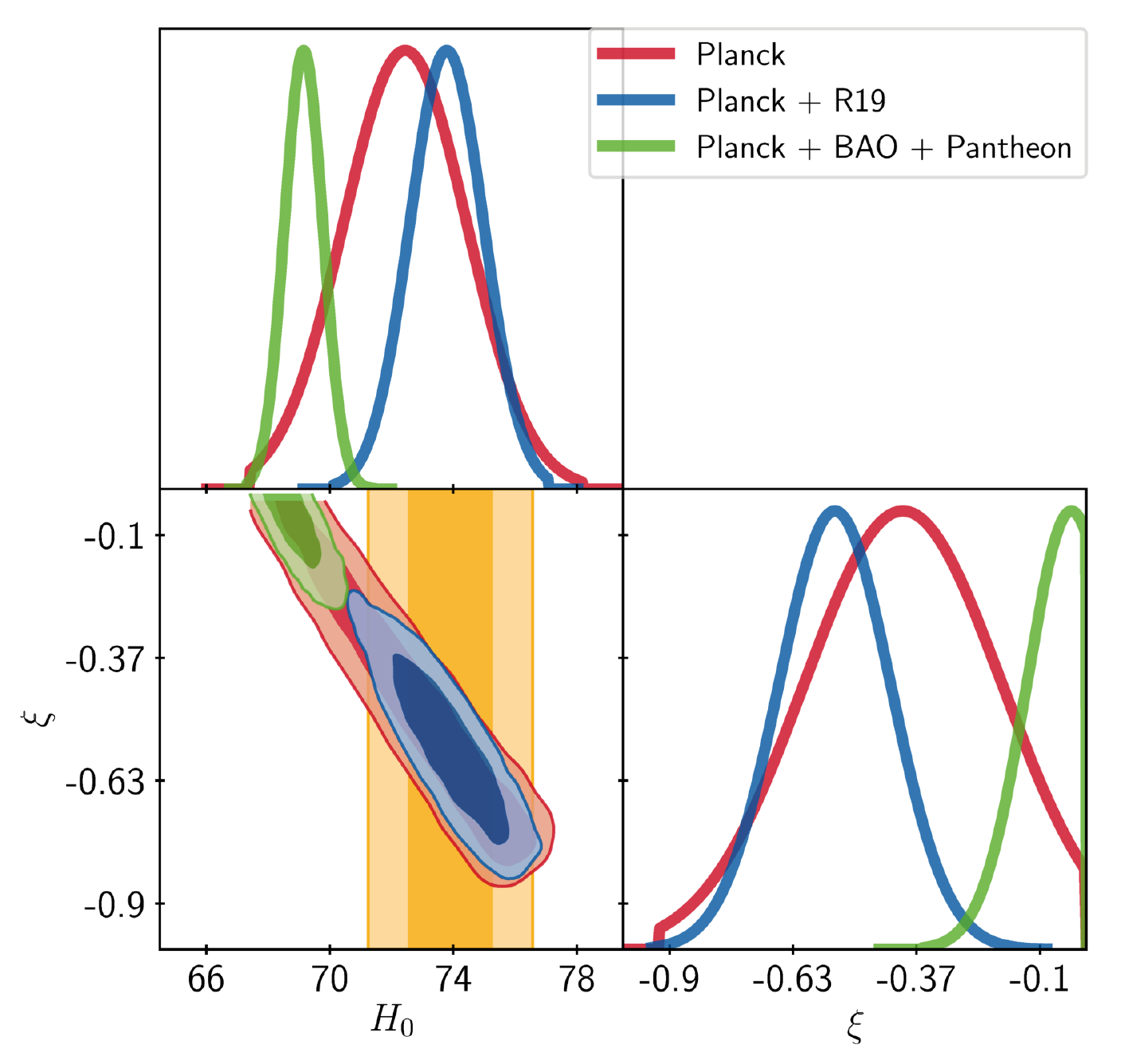}
	\caption{Two-dimensional contours (68\% and 95\%  C.L.) of the ($\xi-H_0$) plane. The different colors denote different combinations of probes considered within this work: \textit{Planck} (red), \textit{Planck}+R19 (blue), and \textit{Planck}+BAO+Pantheon (green). The yellow band corresponds to the R19 measurement.}
	\label{fig:MCMC_result}
\end{figure}

Here we present an overview of the current cosmological constraints on iDMDE. A summary of the parameters most significantly affected by iDMDE is presented in Table~\ref{tab:bestfits}, where each column refers to a given combination of data sets. A similar set of detectors can be found in Table 1 of \cite{DiValentino:2019jae}, with different combinations. Here we emphasize the important role of combining these different data sets, especially BAO and Pantheon. Furthermore, we also show our most relevant results in Fig.~\ref{fig:MCMC_result}.

The first two cases, \textit{Planck} and \textit{Planck}+R19, can be compared to the work already presented in e.g.~\cite{DiValentino:2019ffd}. As already suggested there, this combination allows iDMDE to reconcile the \textit{Planck} predictions with the late-time R19 measurements, yielding a value of ${H_0=74.0_{-1.3}^{+1.4}~\text{km/(s Mpc)}}$. However, when extending the analysis to the combination of \textit{Planck}+BAO+Pan-theon, the preference for a higher $H_0$ value is substantially mitigated.

Indeed, in Fig.~\ref{fig:MCMC_result} we can see that the \textit{Planck}+R19 and the \textit{Planck}+BAO+Pantheon contours do not overlap at the $2\sigma$ level. This indicates that the preference for a higher $H_0$ value is driven entirely by the inclusion of the R19 data, while Pantheon and BAO data favor a lower value of $H_0=68.78_{-0.74}^{+0.54}$~km/(s Mpc), which is $1.5\sigma$ different to the standard $\Lambda$CDM value from \cite{Aghanim:2018eyx}, and $3.3\sigma$ different to the R19 value. If we consider all data sets together,\footnote{Given that the data sets do not overlap at the $2\sigma$ level, any interpretation of their combination should be taken with great care.} the BAO data lends more weight, leading to a final value of $H_0 = 69.82_{-0.76}^{+0.63}$~km/(s Mpc), which is $2.5\sigma$ from the standard $\Lambda$CDM value, and $2.6\sigma$ from the R19 value. As such, it seems that iDMDE does not allow to fully reconcile the different data sets considered here, but it can considerably reduce the significance of the tension.

\begin{figure}[t]
	\centering
	\includegraphics[width=\columnwidth]{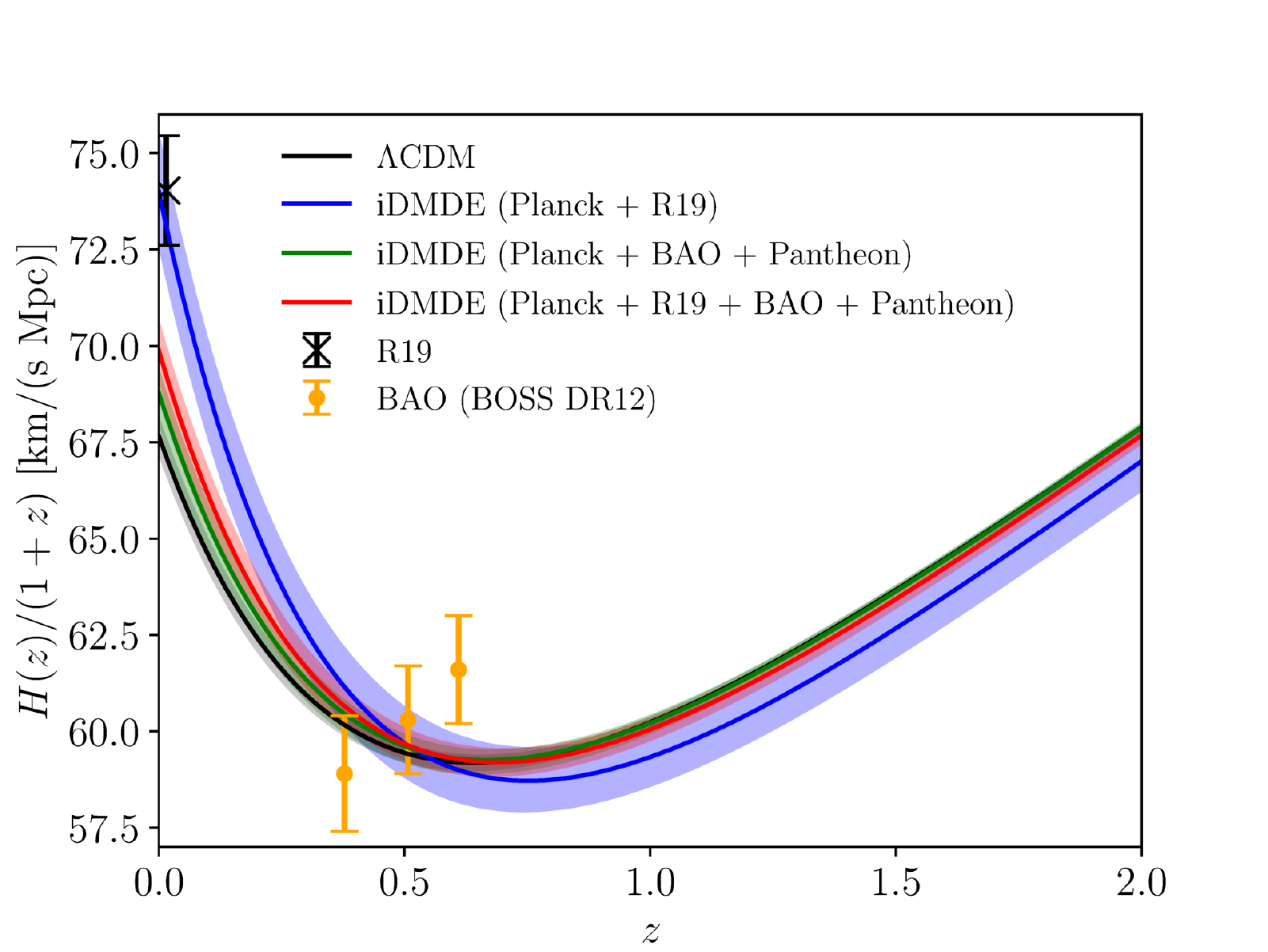}
	\caption{Evolution of $H(z)$ for iDMDE, using the best-fit cosmological parameters from Table \ref{tab:bestfits} for \textit{Planck}+R19 (blue), \textit{Planck}+ BAO+Pantheon (green), and \textit{Planck}+R19+BAO+ Pantheon (red). For comparison, the standard $\Lambda$CDM prediction is also shown in black. The shaded areas correspond to the 1$\sigma$ bounds. Additionally, the R19 data point is shown, as well as several low-redshift BAO measurements.}
	\label{fig:H0_curves}
\end{figure}

\begin{table*}
	\centering
	\begin{tabular}{  c@{\hskip 0.8 cm}  ccc@{\hskip 0.8 cm}  ccc@{\hskip 0.8 cm}  ccc }
		\hline\rule{0pt}{3.0ex} 
		& \multicolumn{3}{c}{\thead{\textit{Planck} \\ + R19 }} & \multicolumn{3}{c}{\thead{\textit{Planck} \\ + BAO \\+ Pantheon}} & \multicolumn{3}{c}{\thead{\textit{Planck} \\ + R19 \\ + BAO \\+ Pantheon}} \\ [0.3 cm]
		\hline\rule{0pt}{3.0ex}		
		&  $\Lambda$CDM & iDMDE & $\Delta \chi^2$ & $\Lambda$CDM & iDMDE & $\Delta \chi^2$ & $\Lambda$CDM & iDMDE & $\Delta \chi^2$\\[0.1 cm]
		\hline\rule{0pt}{3.0ex} 
		\textit{Planck} high-$\ell$ TTTEEE &$ 2350.39 $&$ 2346.24 $&$  -4.15 $&$ 2346.57 $&$ 2347.11 $&$  0.54 $&$ 2348.39 $&$ 2349.59 $&$   1.20 $\\
		\textit{Planck} low-$\ell$ TT      &$   22.84 $&$   23.85 $&$   1.01 $&$   23.57 $&$   23.75 $&$  0.18 $&$   23.31 $&$   22.98 $&$  -0.33 $\\
		\textit{Planck} low-$\ell$ EE      &$  397.86 $&$  395.83 $&$  -2.03 $&$  396.77 $&$  395.69 $&$ -1.08 $&$  395.74 $&$  396.17 $&$   0.43 $\\
		\textit{Planck} lensing 			  &$    9.17 $&$    8.79 $&$  -0.38 $&$    8.81 $&$    8.87 $&$  0.06 $&$    9.27 $&$    8.84 $&$  -0.43 $\\
		R19                       &$   12.06 $&$    0.04 $&$ -12.02 $&$       - $&$       - $&$     - $&$   15.24 $&$   10.43 $&$  -4.81 $\\
		Pantheon                  &$       - $&$       - $&$      - $&$ 1027.08 $&$ 1027.04 $&$ -0.04 $&$ 1027.20 $&$ 1027.43 $&$   0.23 $\\
		BAO                       &$       - $&$       - $&$      - $&$    5.17 $&$    5.35 $&$  0.18 $&$     5.3 $&$    6.86 $&$   1.56 $\\
		\hline\rule{0pt}{3.0ex} 
		Total                     &$ 2792.33 $&$ 2774.75 $&$ -17.58 $&$ 3807.97 $&$ 3807.83 $&$ -0.14 $&$ 3824.45 $&$ 3822.30 $&$  -2.15 $\\[0.1 cm]
		\hline
	\end{tabular}
	\caption{Comparison of $\Lambda$CDM and iDMDE, showing the $\chi^2$ contribution from each individual data set, for three different runs. A negative $\Delta \chi^2$ indicates a preference for iDMDE, while a positive $\Delta \chi^2$ indicates a preference for $\Lambda$CDM.}
	\label{tab:chi2}
\end{table*}

This is further illustrated in Fig.~\ref{fig:H0_curves}, where we show the late-time evolution of $H(z)$ for iDMDE, using the best fits obtained for the different data combinations from Table~\ref{tab:bestfits}. For comparison, the evolution of $H(z)$ within $\Lambda$CDM is also shown, using the best fits from the last column of Table 2 of~\cite{Aghanim:2018eyx}. When using \textit{Planck}+R19, we are able to account for both the early-time $H(z_\text{CMB})$ and the late-time $H(z_\text{R19})$, thus bringing these data sets into closer agreement than in $\Lambda$CDM. However, when using BAO and Pantheon data, the former drive $H(z)$ to lower values today, no longer fully solving the $H_0$ tension.

The results presented here are consistent with previous analyses on late-time solutions to the Hubble tension (see e.g.~\cite{Bernal:2016gxb, Poulin:2018zxs, Knox:2019rjx}). Indeed our Fig.~\ref{fig:H0_curves} can be compared to Fig.~8 of~\cite{Bernal:2016gxb} or Fig. 3 of~\cite{Poulin:2018zxs}. This shows that, despite the presence of the additional interacting terms, the iDMDE model cannot avoid the \emph{no-go theorem} found for late-time modifications to the expansion history of the universe.

Moreover, in addition to the implications for the Hubble tension discussed above, several interesting conclusions on the ability of the iDMDE model to reconcile the different probes can be drawn by performing several statistical analyses. First, we can see that the $\Delta\chi^2$ given in Table~\ref{tab:bestfits} are negative for all data set combinations, indicating a (mostly mild) preference for \text{iDMDE} over $\Lambda$CDM. For the cases of \textit{Planck} alone and \textit{Planck}+BAO+Pantheon, the improvement is not statistically significant when taking into consideration the addition of the free parameter $\xi$, increasing the degrees of freedom by one, as can be seen also by looking at the $\sigma$ values (1.9 and 0.4 respectively). However, the inclusion of R19 data substantially increases the preference
for \text{iDMDE} over $\Lambda$CDM, with $\Delta\chi^2 = -17.58$ (corresponding to  $\sim4.2\sigma$). The impact of each additional likelihood on the total $\chi^2$ is explored in detail in Table~\ref{tab:chi2}, where we can see the biggest contribution to the negative $\Delta\chi^2$ is from R19. Finally, as show in the last columns of Tabs.~\ref{tab:bestfits} and~\ref{tab:chi2}, when considering all data sets together, we find a $\Delta\chi^2$ of $-2.15$ with the inclusion of one additional free parameter (corresponding to a $\sim 1.5\sigma$ preference).

Furthermore, we can use the Bayes ratio defined in Eq.~\eqref{eq:bayes} to see that the \textit{Planck} and \textit{Planck}+BAO+\linebreak Pantheon data sets show a positive preference for $\Lambda$CDM. On the other hand, the combination of \textit{Planck}+R19 shows a strong preference for iDMDE, due to its ability to reconcile these two data sets. However, when considering all data sets together, the Bayes ratio indicates only a very mild preference for iDMDE. Thus, we conclude that there is no clear preference for the iDMDE model considered within this work over $\Lambda$CDM.

Finally, note that, although not quantitatively shown in this work, the behavior we described in this section can also be observed in models where the DE EOS parameter $w$ is left as a free parameter (see e.g. Fig. 2 of~\cite{DiValentino:2019jae}). Furthermore, the results obtained for extensions of the energy transfer function expressed in Eq. \eqref{eq: Q}, such as those considered in \cite{Pan:2020bur}, hint to the same conclusion found in this work, with similar tensions among the different data sets, although less pronounced (see e.g. Fig.~8 of the reference).

\section{Conclusions}\label{sec: conc}
With the increasing level of precision obtained by CMB and local distance ladder measurements missions, such as \textit{Planck} and HST, as well as by cosmic shear surveys, such as KiDS and DES, we have seen the rise of significant tensions in the cosmological landscape. One such tension that has gained a lot of attention is the 4.4$\sigma$ discrepancy between the values of the expansion rate of the universe, $H_0$, as reported by the \textit{Planck} and SH0ES collaborations.

In order to address this tension, a variety of different models have been proposed. Within this work, we focused on a class of models which allows for interactions between DM and DE. Specifically, we considered the possibility that a coupling term linking the energy density conservation equations for these two fluids is present, and is linear in the DE energy density. Furthermore, we assume a flat potential for the DE fluid, which differentiates this model from other existing coupled DE scenarios.

This scenario has already been very well studied in the literature due to its potential to alleviate the $H_0$ tension. In fact, as shown in the literature as well as in this work, when considering the combination of \textit{Planck}+R19 data, the model allows for significantly higher $H_0$ values than those predicted by $\Lambda$CDM. However, we have shown here that when considering \textit{Planck}+BAO+Pantheon, this preference for a higher $H_0$ value is substantially mitigated, leading to $H_0=68.78_{-0.74}^{+0.54}$~km/(s Mpc), which is within $1.5\sigma$ of the standard $\Lambda$CDM value from \cite{Aghanim:2018eyx}, and $3.3\sigma$ away from the R19 value. As such, we find that while the model can slightly alleviate the $H_0$ tension, it is not able to conclusively solve it.

Furthermore, when all aforementioned cosmological probes are considered together, we find the preferred value of the Hubble parameter to be ${H_0 = 69.82_{-0.76}^{+0.63}}$~km/(s Mpc), which is $2.5\sigma$ form the standard $\Lambda$CDM value and $2.6\sigma$ from the latest local measurements. Moreover, for this combination of data sets, the detailed $\chi^2$ analysis performed in this work yields only a $\Delta\chi^2$ of $-2.15$ when compared to the base $\Lambda$CDM, with the inclusion of one additional free parameter (corresponding to $\sim 1.5\sigma$). Additionally, an analysis of the Bayes ratio finds no strong preference for either model.

Thus, we conclude that, although the iDMDE model considered in this work can significantly alleviate the Hubble tension, the data shows no statistical preference for it over $\Lambda$CDM.

\section*{Acknowledgements}
We thank Thomas Hambye, Julien Lesgourgues, Laura Lopez-Honorez, Sunny Vagnozzi, and Matteo Viel for very useful discussions. ML is supported by the \tquote{Probing dark matter with neutrinos} ULB-ARC convention. DH is supported by the FNRS research grant number~\mbox{F.4520.19}.

\bibliography{biblio}{}

\end{document}